\documentclass[11pt]{article}
\usepackage[letterpaper,margin=1in]{geometry}
\usepackage{amsmath,amssymb,amsthm}

\usepackage[T1]{fontenc}
\usepackage{comment}
\usepackage{lmodern,microtype}
\usepackage{xcolor}
\usepackage[colorlinks=true,allcolors=blue!45!black]{hyperref}
\hypersetup{pdftitle={A short proof of the AA simulation conjecture via the completion framework}}

\usepackage{mathtools}
\usepackage[capitalise,noabbrev,nameinlink]{cleveref}

\newtheorem{theorem}{Theorem}
\newtheorem{lemma}[theorem]{Lemma}

\newtheorem{conjecture}[theorem]{Conjecture}

\crefname{conjecture}{Conjecture}{Conjectures}
\Crefname{conjecture}{Conjecture}{Conjectures}
\crefname{equation}{Eq.}{Eqs.}
\Crefname{equation}{Eq.}{Eqs.}
\newcommand{\cube}{\{0,1\}^n}
\newcommand{\goodset}{\mathcal{D}}
\newcommand{\E}{\mathbb{E}}
\newcommand{\R}{\mathbb{R}}
\newcommand{\one}{\mathbf{1}}
\newcommand{\eps}{\varepsilon}
\newcommand{\density}{\delta}
\newcommand{\certbound}{\gamma}
\newcommand{\diff}{\Delta}
\DeclarePairedDelimiter{\abs}{\lvert}{\rvert}
\DeclarePairedDelimiter{\norm}{\lVert}{\rVert}
\newcommand{\cost}[1]{\norm{#1}_1}
\newcommand{\Inf}{I}

\newcommand{\restr}[2]{#1|_{#2}}
\DeclareMathOperator{\D}{D}
\DeclareMathOperator{\Q}{Q}
\DeclareMathOperator{\FC}{FC}

\DeclareMathOperator{\weight}{wt}
\DeclareMathOperator{\CAdv}{CAdv}
\DeclareMathOperator{\poly}{poly}
\allowdisplaybreaks[1]
\DeclareMathOperator{\Var}{Var}
\DeclareMathOperator{\Cov}{Cov}
\AddToHook{env/align/begin}{%
    \setlength{\abovedisplayskip}{8pt plus 2pt minus 2pt}%
    \setlength{\belowdisplayskip}{8pt plus 2pt minus 2pt}%
}
\AddToHook{env/align*/begin}{%
    \setlength{\abovedisplayskip}{8pt plus 2pt minus 2pt}%
    \setlength{\belowdisplayskip}{8pt plus 2pt minus 2pt}%
}
\DeclareMathOperator{\domain}{\mathrm{Dom}}

\title{A short proof of the Aaronson-Ambainis Conjecture\\via the completion framework}
\author{%
{\normalsize
\begin{tabular}{@{}c@{\hspace{2.2em}}c@{\hspace{2.2em}}c@{}}
Thomas Huffstutler & David Miloschewsky & Supartha Podder
\end{tabular}}\\[0.5em]
{\small Computer Science, Stony Brook University, New York, USA}\\[0.2em]
{\footnotesize
\texttt{\{thuffstutler, dmiloschewsk, supartha\}@cs.stonybrook.edu}}
}

\date{\small\today}

\begin{document}
\maketitle
\vspace{-2em}
\begin{abstract}
We give a short proof of the Aaronson-Ambainis conjecture (Conjecture~7.1 of~\cite{AA14}). If a partial Boolean function $f$ is defined on a $\density$-fraction of all $n$-bit strings and have quantum query complexity $Q(f)$, then, for every $0<\eps<1$, a deterministic algorithm computes $f$ correctly on at least a $1-\eps$ fraction of its domain using $O\left(\Q(f)^6 / (\eps\density)^2\right)$ queries. Our proof uses the completion framework of~\cite{HKMP26} and establishes a more general bound in terms of the classical adversary measure. The only result we use from Turturean's recent proof~\cite{Tur26} is a special case of his certificate inequality, for which we provide a self-contained proof.

We also derive the stronger variant of the Aaronson-Ambainis Conjecture, the bounded-polynomial influence conjecture (Conjecture~1.7 of~\cite{AA14}), from our main result in Appendix~\ref{app:polynomial-influence}.
\end{abstract}

\section{The conjecture and our completion theorem}

On November 6th, 2026, Turturean posted a 60-page preprint containing a proof of the Aaronson-Ambainis conjecture~\cite{Tur26}. We give a short proof of the following equivalent version of the Aaronson-Ambainis conjecture. 

\begin{conjecture}[Conjecture 7.1 of~\cite{AA14}]
\label{conj:aa}
    Let $f$ be a partial Boolean function with $\domain(f) \neq \emptyset$ and density $\density=\abs{\domain(f)}/2^n$.
    For every $0<\eps<1$, there is a deterministic algorithm making $\poly(\Q(f),1/\eps,1/\density)$ queries that computes $f$ correctly on at least a $1-\eps$ fraction of $\domain(f)$.
\end{conjecture}

Our approach uses the completion framework of~\cite{HKMP26}. Given a partial Boolean function $f$, a total Boolean function $F$ is a \emph{completion} of $f$ if $F(x)=f(x)$ for all $x \in \domain(f)$, while values on inputs outside the domain are assigned arbitrarily. For the Aaronson-Ambainis conjecture, we only need correctness \emph{on most} of the inputs in $\domain(f)$. Therefore, we allow ourselves to discard at most an $\eps$ fraction of $\domain(f)$ before extending the function. Specifically, we seek $\goodset\subseteq\domain(f)$ with $\abs{\goodset}\geq(1-\eps)\abs{\domain(f)}$ and a total Boolean function $F$ agreeing with $f$ on $\goodset$, together with a deterministic algorithm for $F$ using $\poly(\Q(f),1/\eps,1/\density)$ queries that computes $f$ correctly on every input in $\goodset$, without needing to test membership in $\goodset$.
Conversely, any deterministic algorithm achieving the desired accuracy in computing $f$ naturally defines a total Boolean function $F$, with $\goodset=\{x\in\domain(f):F(x)=f(x)\}$. Thus, constructing a low-complexity completion after discarding a small fraction of the domain is precisely the desired simulation.

We prove a statement in terms of the classical adversary measure $\CAdv$, rather than quantum query complexity.
Write $A_b=f^{-1}(b)$ and $\diff(x,y)=\{i:x_i\neq y_i\}$.
For a nonnegative vector $v$, write $\cost{v}=\sum_i v_i$ for its cost.
Using the formulation of~\cite{ABK21}, $\CAdv(f)$ is the minimum of $\max_{x\in\domain(f)}\cost{\lambda^x}$ over vectors $\lambda^x\in\R_{\geq0}^n$, $x\in\domain(f)$, satisfying
\begin{align}
    \sum_{i\in\diff(x,y)}\min\{\lambda_i^x,\lambda_i^y\} &\geq1 \quad\text{if }f(x)\ne f(y).\label{eq:compatible}
\end{align}

\begin{theorem}[Completion from classical adversary bounds]\label{thm:main}
    Let $f$ be a partial Boolean function with $\domain(f) \neq \emptyset$ and density $\density=\abs{\domain(f)}/2^n$.
    For every $0<\eps<1$, there are $\goodset\subseteq\domain(f)$ and a total function $F:\cube\to\{0,1\}$ such that
    \begin{align*}
        \abs{\goodset} &\geq(1-\eps)\abs{\domain(f)},\qquad F\supseteq\restr{f}{\goodset},\qquad \D(F)=O\left(\frac{\CAdv(f)^3}{(\eps\density)^2}\right).
    \end{align*}
\end{theorem}

\cref{thm:main} with the bound $\CAdv(f)=O(\Q(f)^2)$ of~\cite{ABK21} immediately implies \cref{conj:aa} with $O(\Q(f)^6/(\eps\density)^2)$ queries. 

The proof first constructs a completion $G\supseteq f$ with total influence $\Inf(G)=O(\CAdv(f))$ and fractional certificate complexity $\FC(G,x)=O(\CAdv(f))$ for every $x\in\domain(f)$. At each step of our deterministic algorithm, the inputs consistent with the answers so far form a subcube. We apply \cref{lem:subset} to this subcube to choose an influential coordinate to query next. Each query decreases the expected remaining total influence by the queried coordinate's influence on the current subcube, so \cref{lem:subset} and the bound on $\Inf(G)$ give a deterministic algorithm using $O(\CAdv(f)^3/(\eps\density)^2)$ queries that agrees with $f$ on at least a $1-\eps$ fraction of its domain.

\section{Proof of the main theorem}

Unless otherwise specified, inputs $x,y,z$ are over $\cube$. Write $x^S$ for $x$ with bits of $S\subseteq [n]$ flipped and $x^i = x^{\{i\}}$. A subcube $C$ of $\cube$ is obtained by fixing the values of some coordinates in $[n]$, the remaining are called free. For a total Boolean function $G$ and $X$ uniform over $\cube$, define the influence of $G$ as,
\begin{align*}
    \Inf_i(G) &=\Pr[G(X)\neq G(X^i)],\qquad \Inf(G)=\sum_i \Inf_i(G).
\end{align*}
A fractional certificate for $G$ at $x$ is a vector $c^x\in\R_{\geq0}^n$ satisfying
\begin{align*}
    \sum_{i\in\diff(x,y)}c_i^x &\geq1
    \qquad\text{for every $y\in\cube$ s.t. $G(y)\neq G(x)$.}
\end{align*}
Define $\FC(G,x)\coloneq\min_{c^x}\cost{c^x}$, where the minimum is over all such fractional certificates. On a subcube $C$, these definitions use the free coordinates of $C$ and take $X$ uniformly over $C$. 

\begin{lemma}\label{lem:completion}
    Every nonconstant partial Boolean function $f$ has a completion $G\supseteq f$ such that,
    \begin{align*}
         \Inf(G) &=O(\CAdv(f)),\qquad \FC(G,x)=O(\CAdv(f))\quad(x\in\domain(f)).
    \end{align*}
\end{lemma}

\begin{proof}
    Choose nonnegative vectors $\lambda^x$ satisfying \cref{eq:compatible} with $\cost{\lambda^x}\leq\CAdv(f)$ for every $x\in\domain(f)$.
    For $a\in\domain(f)$, define $d_a(z)=\sum_{i:z_i\neq a_i}\lambda_i^a$.
    For $a\in A_1$, $b\in A_0$, and any $z\in\cube$, \cref{eq:compatible} gives
    \begin{align}
        d_a(z)+d_b(z) &\geq \sum_{i\in\diff(a,b)}\min\{\lambda_i^a,\lambda_i^b\}\geq1.\label{eq:triangle}
    \end{align}
    Define $g(z) =\min\left\{1,\min_{a\in A_1}d_a(z)\right\}$. Then $g=0$ on $A_1$ and $g=1$ on $A_0$. If $g(z) < 1$, choose $a\in A_1$ such that $d_a(z)=g(z)$.
    For every coordinate $i$, this choice gives
    \begin{align*}
        \max\{0, g(z^i)-g(z)\} &\leq \max\{0, d_a(z^i)-d_a(z)\} \leq \lambda_i^a.
    \end{align*}
    The left-hand sides sum to at most $\CAdv(f)$. When $g(z)=1$, each maximum is $0$. Since $(X,X^i)$ and $(X^i,X)$ have the same distribution,
    \begin{align*}
        L(g) &\coloneq \sum_i\E[\abs{g(X)-g(X^i)}]=2\E\left[\sum_i\max\{0, g(X^i)-g(X)\}\right]=O(\CAdv(f)).
    \end{align*}
    For every threshold $0 < t <1$, define $G_t(x) \coloneq \one[g(x) < t]$. Notice that $G_t$ extends $f$ and $\int_0^1 \Inf(G_t)\,dt= \sum_i\E[\abs{g(X)-g(X^i)}] = L(g)$. Hence, by averaging, there is some $t\in [1/4,3/4]$ with $\Inf(G_t) \leq 2L(g) =O(\CAdv(f))$. Fix the completion $G=G_t$.

    Let us show that $4 \lambda^x$ is a fractional certificate for all $x\in\domain(f)$ of $G$, which implies the statement as $\cost{4 \lambda^x}\leq 4 \CAdv(f)$. For $a\in A_1$, every $y$ such that $G(y)=0$ satisfies $d_a(y) \geq g(y) \geq t \geq 1/4$. Otherwise for $b\in A_0$, every $y$ with $G(y)=1$ has some $a\in A_1$ with $d_a(y)<t$, so \cref{eq:triangle} gives $d_b(y)>1-t\geq1/4$. Thus $\sum_{i\in\diff(x,y)}4\lambda_i^x=4d_x(y)\geq1$ whenever $G(y)\neq G(x)$.
\end{proof}

Note that \cref{lem:completion} only controls certificates on the domain $\domain(f)$. Next we give a self-contained proof of a consequence of Theorem 5.4 in~\cite{Tur26}.

\begin{lemma}\label{lem:subset}
    For $G:\cube \to \{0,1\}$ and $b\in\{0,1\}$, let $U_{b}\subseteq G^{-1}(b)$ be nonempty with density $\density_{b}=\abs{U_{b}}/2^{n}$.
    If $\FC(G,x)\leq \certbound_{b}$ for every $x\in U_b$, then some $i\in [n]$ satisfies
    \begin{align*} 
        \Inf_{i}(G)&\geq\frac{\density_{0}\density_{1}}{\certbound_{0}\certbound_{1}}.
    \end{align*}
    The same holds on a subcube, with densities measured within it.
\end{lemma}
\begin{proof}
    For each $x\in U_b$, fix some fractional certificate $c^x$ such that $\cost{c^x} \leq \certbound_b$. For $i\in [n]$ and $S\subseteq [n]$, define $c^x(S) \coloneq \sum_{i\in S} c^x_i$ and the weights
    \begin{align*}
        w_i&\coloneq\frac{2 \certbound_{0}}{\abs{U_1}}\sum_{x\in U_1}c_i^x, &w(S)&\coloneq\sum_{i\in S}w_i.
    \end{align*}
    Notice that $\cost{w} \leq 2 \certbound_{0}\certbound_{1}$. For $y\in U_0$, let $H_y=\{i:c_i^y>w_i\}$. Since $w(H_y)\leq c^y(H_y)\leq \certbound_{0}$, $\E_{x\in U_1}[c^x(H_y)]=w(H_y)/(2\certbound_{0})\leq 1/2$. By Markov's inequality, at least half of the inputs $x\in U_1$ satisfy $c^x(H_y)<1$. Mark each such pair $(x,y)$. If $(x,y)$ is marked, then $G(x) = G(x^{S})$ for every $S\subseteq H_y$ since $c^x(H_y) < 1$. Furthermore, if $G(z)=1$ and $z$ agrees with $y$ on $H_y$, then
    \begin{align}
        1 \leq c^y(\diff(y,z)) \leq w(\diff(y,z)).\label{eq:certificate-weight}
    \end{align}

    Fix a subcube $C$ with free coordinates $J$ and let $X_b=\{x\in C:G(x)=b\}$ for $b\in\{0,1\}$. Let $N_C$ count the marked pairs $(x,y)$ in $C$ such that $x^J = y$. For $i\in J$, let $b_i$ count the pairs $\{x,x^i\}$ with one input in $X_0$ and the other in $X_1$. Our goal is to prove
    \begin{align}
        N_C &\leq \sum_{i\in J}w_i b_i .\label{eq:subcube-count}
    \end{align}
    When $X_1 = \emptyset$, this is trivial. Otherwise, for $i\in J$ let $t_i$ be variables and define the $\abs{X_1} \times \abs{X_1}$ matrix $M$ by
    \begin{align}
        M_{xz}&=\prod_{i\in J:\,x_i=z_i}t_i\qquad(x,z\in X_1).\label{eq:matrix_def}
    \end{align}
    Notice that $\det(M) \neq 0$ because the product of the diagonal entries has degree $\abs{J} \abs{X_1}$, while every other product in the determinant expansion has smaller degree since each contains an off-diagonal entry, which omits at least one factor $t_i$.

    Fix $x,z\in X_1$ and $K\subseteq J$ such that $x^S \in X_1$ for every $S\subseteq K$. To find factors of $\det(M)$, we use the following identity,
    \begin{align}
        \sum_{S\subseteq K}(-1)^{\abs{S}}\left(\prod_{i\in S}t_i\right)M_{x^S,z}&=\one[K\subseteq\diff(x,z)]\,M_{xz}\prod_{i\in K}(1-t_i^2).\label{eq:coordinate-cancel}
    \end{align}
    Each coordinate $i\in K$ contributes $t_i-t_i=0$ when $x_i=z_i$ and $1-t_i^2$ when $x_i\neq z_i$. For a monomial, define its weight as $\weight\left(\prod_{i\in J} t_i^{\beta_i}\right) = \sum_{i\in J} w_i \beta_i$. Fix a monomial $ m = \prod_{i\in J}t_i^{\beta_i}$ of smallest degree with nonzero coefficient in $\det(M)$. We will show $N_C \leq \weight(m) = \sum_{i\in J}w_i\beta_i \leq\sum_{i\in J}w_i b_i$, which implies \cref{eq:subcube-count}.

    Fix $i\in J$ and let $r_i$ count the disjoint pairs $\{x,x^i\}\subseteq X_1$. Notice that $\abs{X_1} = 2r_i + b_i$. For each such pair, choose either input as $x$ and over all $z\in X_1$, set $M_{xz}$ to $M_{xz}-t_i M_{x^i,z}$.
    These operations preserve $\det(M)$ and by \cref{eq:coordinate-cancel} with $K=\{i\}$ it inserts a factor $(1-t_i^2)$ in each new row. Therefore there is a polynomial $P_i$ such that $\det(M) = (1-t_i^2)^{r_i} P_i$ whose degree in $t_i$ is $\deg_{t_i}(P_i) = \deg_{t_i}(\det(M)) - 2r_i \leq \abs{X_1} - 2 r_i = b_i$. By the expansion of $\det(M)$ using $P_i$, the least-degree terms come from $P_i$ because the first term has the term $1$. Therefore, $\beta_i = \deg_{t_i}(m) \leq \deg_{t_i}(P_i) \leq b_i$, implying that $\sum_{i \in J} w_i \beta_i \leq \sum_{i \in J} w_i b_i$.

    Next we show that $N_C \leq \weight(m)$. Starting with the original matrix $M$, we transform it into $\widetilde{M}$ as follows. For each pair $(x,y)$ counted by $N_C$, let $K=H_y \cap J$ and update the row $x$ by setting $\widetilde{M}_{x,z}$ to the left side in \cref{eq:coordinate-cancel} over all $z\in X_1$. Note that the rows $x^S$ exist for every $S\subseteq K$ as $(x,y)$ is marked. The rest of $\widetilde{M}$ is as in $M$. These replacements affect $N_C$ distinct rows because $y=x^J$ is determined by $x$.

    By linearity in each row, we expand $\det(\widetilde{M})$ in terms of the original rows using the sums over $S$ in \cref{eq:coordinate-cancel}. When $S=\emptyset$ in every replacement, we have $\det(M)$. If a row repeats, the contribution is zero. Otherwise, it is $\pm\det(M)$ multiplied by the factors from the row replacements. Unless every $S=\emptyset$, these factors have positive degree so the contribution cannot change the coefficient of $m$.

    Consider a pair $(x,y)$ counted by $N_C$ and a column $z\in X_1$ with $\widetilde{M}_{x,z} \neq 0$. By \cref{eq:coordinate-cancel}, $z$ differs from $x$ on $K$, so $z$ agrees with $y$ there. As they also agree outside of $J$, they agree on all of $H_y$. Therefore \cref{eq:matrix_def,eq:certificate-weight} gives $M_{xz} = \prod_{i\in\diff(y,z)} t_i$ and $w(\diff(y,z))\geq 1$. Expanding the factors in \cref{eq:coordinate-cancel}, each monomial in this entry is up to sign $M_{xz}\prod_{i\in S}t_i^2$ for some $S\subseteq K$ and
    \begin{align*}
        \weight \left(M_{xz}\prod_{i\in S}t_i^2\right) =w(\diff(y,z))+2\sum_{i\in S} w_i\geq1.
    \end{align*}
    Every product in the determinant expansion uses one entry from each of the $N_C$ replaced rows, while all other entries have nonnegative weight. Hence every monomial in $\det(\widetilde M)$ has weight at least $N_C$. Since $m$ occurs in $\det(\widetilde M)$, we obtain $N_C \leq\weight(m)$, completing the proof of \cref{eq:subcube-count}.
    
    Next we sum over all subcubes. Each marked pair $(x,y)$ is counted by $N_C$ for exactly one subcube $C$ whose free coordinates are $\diff(x,y)$, while each $\{x,x^i\}$ with $G(x)\neq G(x^i)$ belongs to $2^{n-1}$ subcubes, since every other coordinate can be fixed or free. There are $2^{n-1}\Inf_i(G)$ such pairs. Summing \cref{eq:subcube-count} and using $\cost{w}\leq2\certbound_{0}\certbound_{1}$ gives
    \begin{align*}
        \frac{\abs{U_0}\,\abs{U_1}}{2}&\leq 4^{n-1}\sum_i w_i \Inf_i(G)\leq 2\cdot4^{n-1}\certbound_{0}\certbound_{1}\max_i \Inf_i(G).
    \end{align*}
    Substituting $\abs{U_b}=2^n\density_b$ proves the bound.
\end{proof}

The lemmas allow us to prove the main result.

\begin{proof}[Proof of \cref{thm:main}]
    If $f$ is constant, let its completion $F$ also be constant. Otherwise choose $G$ as in \cref{lem:completion}.
    For a subcube $C$, define
    \begin{align*}
        \density_b(C) &=\Pr[X\in A_b\mid X\in C]=\frac{\abs{A_b\cap C}}{\abs{C}},\qquad m(C)=\min\{\density_0(C),\density_1(C)\}.
    \end{align*}
    The certificates at inputs in $\domain(f)\cap C$ remain valid after restricting to the free coordinates of $C$.
    If $m(C)>0$, \cref{lem:subset} applied with $U_b=A_b\cap C$ implies a free coordinate $i$ with
    \begin{align}
        \Inf_i(\restr{G}{C}) &=\Omega\left(\frac{m(C)^2}{\CAdv(f)^2}\right),\label{eq:drop}
    \end{align}
    as $\density_0(C)\density_1(C)\geq m(C)^2$. The algorithm repeatedly queries such $i$ until $m(C)=0$. Next, we track the expected remaining total influence.
    For $C_b=C\cap\{x_i=b\}$, the exact identity
    \begin{align}
        \Inf(\restr{G}{C})-\frac{\Inf(\restr{G}{C_0})+\Inf(\restr{G}{C_1})}2 &=\Inf_i(\restr{G}{C})\label{eq:influence-split}
    \end{align}
    holds because all other coordinate influences average unchanged. For $t\geq 0$, run the procedure until $m(C)=0$ or $t$ queries have been made. When done, output the more frequent value of $f$ among the inputs in $\domain(f) \cap C$, choosing either in case of a tie. Let $X$ be a uniformly random input from $\cube$ and $C_t(X)$ the subcube consistent with the answers obtained on $X$. Then $e_t\coloneq\E[m(C_t(X))]$ is the fraction of all inputs in $\cube$ that belong to $\domain(f)$ and the procedure outputs incorrectly. Additional queries cannot increase $e_t$. Telescoping~\cref{eq:influence-split}, using~\cref{eq:drop}, and applying Jensen's inequality gives for $s\geq1$,
    \begin{align*}
        s e_s^2 &\leq\sum_{t=0}^{s-1}e_t^2\leq\sum_{t=0}^{s-1}\E[m(C_t(X))^2]=O(\CAdv(f)^2 \Inf(G))=O(\CAdv(f)^3).
    \end{align*}
    Hence $e_s=O(\sqrt{\CAdv(f)^3/s})$. Choose $s=O(\CAdv(f)^3/(\eps\density)^2)$ be sufficiently large so that $e_s\leq\eps\density$. The algorithm then computes a total function $F$ using at most $s$ queries. For $\goodset=\{x\in\domain(f):F(x)=f(x)\}$, we have
    \begin{align*}
        \abs{\domain(f)\setminus\goodset} &\leq\eps\density\,2^n=\eps\abs{\domain(f)}.\qedhere
    \end{align*}
\end{proof}

\paragraph{AI Disclosure.} We have long believed that the most promising route to the Aaronson-Ambainis conjecture was through its equivalent formulation in Conjecture~7.1 of~\cite{AA14}, with our completion framework providing a natural approach. Indeed, we formulated the conjecture as a completion question in~\cite{HKMP26}.

Over the four months preceding Turturean's result, we explored this route extensively with the assistance of ChatGPT 5.6 and 6. These efforts led to ideas similar to those used in the present proof, but we were unable to obtain an ingredient playing the role of Turturean's certificate inequality~\cite{Tur26}. His result supplied the missing ingredient that allowed this approach to succeed. We use a special case of his certificate inequality, which we reprove here in a self-contained form for completeness.

\paragraph{Acknowledgment.}
We thank Ashley Montanaro for asking whether our argument also proves the bounded-polynomial influence conjecture (Conjecture~1.7 of~\cite{AA14}), which led to the extension presented in the appendix.
%After the release of our five-page draft, Ashley Montanaro asked whether our argument also proves the bounded-polynomial influence conjecture. We thank him for this question, which led to the extension presented in the appendix.

\bibliographystyle{alpha}
\bibliography{bibliography}

\begin{thebibliography}{HKMP26}

\bibitem[AA14]{AA14}
Scott Aaronson and Andris Ambainis.
\newblock The need for structure in quantum speedups.
\newblock {\em Theory of Computing}, 10(6):133--166, 2014.
\newblock DOI:
  \href{https://doi.org/10.4086/toc.2014.v010a006}{10.4086/toc.2014.v010a006}.

\bibitem[ABDK21]{ABK21}
Anurag Anshu, Shalev Ben-David, and Srijita Kundu.
\newblock On query-to-communication lifting for adversary bounds.
\newblock In {\em CCC 2021}, volume 200 of {\em LIPIcs}, pages 30:1--30:39,
  2021.
\newblock DOI:
  \href{https://doi.org/10.4230/LIPIcs.CCC.2021.30}{10.4230/LIPIcs.CCC.2021.30}.

\bibitem[HKMP26]{HKMP26}
Thomas Huffstutler, Upendra Kapshikar, David Miloschewsky, and Supartha Podder.
\newblock {A Framework for Ruling out Quantum Speedups}.
\newblock In {\em MFCS 2026}, volume 386 of {\em LIPIcs}, pages 10:1--10:17,
  2026.
\newblock DOI:
  \href{https://doi.org/10.4230/LIPIcs.MFCS.2026.10}{10.4230/LIPIcs.MFCS.2026.10}.
  Full version: \href{https://arxiv.org/abs/2603.29256}{arXiv:2603.29256}.

\bibitem[OSSS05]{OSSS05}
Ryan O'Donnell, Michael Saks, Oded Schramm, and Rocco~A Servedio.
\newblock Every decision tree has an influential variable.
\newblock In {\em FOCS 2005}, pages 31--39, 2005.
\newblock DOI:
  \href{https://doi.org/10.1109/SFCS.2005.34}{10.1109/SFCS.2005.34}.

\bibitem[Tur26]{Tur26}
David Turturean.
\newblock A proof of the aaronson-ambainis conjecture, 2026.
\newblock \href{https://arxiv.org/abs/2611.09773}{arXiv:2611.09773}.

\end{thebibliography}

\appendix

\section{The bounded-polynomial influence conjecture}
\label{app:polynomial-influence}

We show that \cref{thm:main} also implies the stronger version of the Aaronson-Ambainis conjecture, the bounded-polynomial influence conjecture.

\begin{conjecture}[Conjecture~1.7 of~\cite{AA14}]\label{cor:aa-influence}
Let $p:\mathbb{R}^n\to\mathbb{R}$ be a polynomial of degree at most
$d\geq1$ satisfying $0\leq p(x)\leq1$ for every $x\in\{0,1\}^n$.
If
\begin{align*}
    \E_{X\in\cube}[(p(X)-\E[p(X)])^2] &\geq\eps>0,
\end{align*}
then there exists some $i\in[n]$ that satisfies
\begin{align*}
        \Inf_i(p) \geq\left(\frac{\eps}{d}\right)^{O(1)}.
    \end{align*}
\end{conjecture}

The proof applies \cref{thm:main} to a partial function defined by $p$, then uses a covariance inequality described in \cref{lem:app-covariance}.

All expectations below are with respect to the uniform measure on $\cube$. For real-valued functions $p$ and $h$, write
\begin{align*}
    \Inf_i(p) &=\E[(p(X)-p(X^i))^2],\qquad \Var(p)=\E[(p-\E[p])^2],\qquad \Cov(p,h)=\E[(p-\E[p])h].
\end{align*}

\begin{lemma}[Corollary of~\cite{OSSS05}]\label{lem:app-covariance}
    Let $p:\cube\to\R$ and $h:\cube\to[-1,1]$. If a deterministic algorithm  computes $h$ using at most $D$ queries, then
    \begin{align*}
        \abs{\Cov(p,h)} &\leq\frac D2\sqrt{\max_i\Inf_i(p)}.
    \end{align*}
\end{lemma}
\begin{proof}
    Let $q_i$ be the probability that the algorithm queries coordinate $i$ on a uniformly random input. In our notation, Eq.~(6) of~\cite{OSSS05} gives
    \begin{align*}
        \abs{\Cov(p,h)} &\leq\frac12\sum_iq_i\E[\abs{p(X)-p(X^i)}].
    \end{align*}
    The factor $1/2$ arises because resampling a uniform bit flips it with probability $1/2$. By Cauchy--Schwarz and $\sum_iq_i\leq D$,
    \begin{align*}
        \abs{\Cov(p,h)} &\leq\frac12\sum_iq_i\sqrt{\Inf_i(p)}
        \leq\frac D2\sqrt{\max_i\Inf_i(p)}.\qedhere
    \end{align*}
\end{proof}

\begin{theorem}\label{thm:app-polynomial-influence}
    Let $n,d\geq1$, and let $p:\cube\to[0,1]$ be a real polynomial of degree $d$. If $v=\Var(p)>0$, then
    \begin{align*}
        \max_{i\in[n]}\Inf_i(p) &= \Omega\left(\frac{v^{12}}{d^{12}}\right)
    \end{align*}
\end{theorem}
\begin{proof}
    Let $z=p-\E[p]$ and define
    \begin{align*}
        \domain(f) &=\{x:\abs{z(x)}\geq v/4\},\qquad f(x)=\one[z(x)>0]\quad(x\in\domain(f)),\qquad \density=\frac{\abs{\domain(f)}}{2^n}.
    \end{align*}
    Since $\abs{z}\leq1$ and $\abs{z}<v/4$ outside $\domain(f)$,
    \begin{align*}
        v=\E[z^2] &\leq\E[\abs{z}]\leq\density+v/4.
    \end{align*}
    Thus $\density\geq3v/4$. The polynomial $(1+z)/2$ has degree at most $d$, is bounded in $[0,1]$ on the entire cube, and approximates $f$ on $\domain(f)$ with error at most $1/2-v/8<1/2$. The error-dependent approximate-degree bound from Theorem 8 of~\cite{ABK21} gives
    \begin{align*}
        \CAdv(f) &\leq\frac{4\pi^2d^2}{v}.
    \end{align*}
    Apply \cref{thm:main} with $\eps=v/(16\density)\in(0,1)$.
    We obtain a total Boolean function $F$ with
    \begin{align*}
        \Pr[X\in\domain(f),\ F(X)\ne f(X)] &\leq v/16
    \end{align*}
    and a deterministic algorithm computing $F$ using at most $D$ queries, where
    \begin{align*}
        D &=O(\CAdv(f)^3/v^2)=O(d^6/v^5).
    \end{align*}
    Set $h=2F-1$ and $E=\{x\in\domain(f):F(x)\ne f(x)\}$.
    On $\domain(f)\setminus E$, we have $zh=\abs{z}$; elsewhere $zh\geq-\abs{z}$.
    Since $\abs{z}<v/4$ on $\domain(f)^c$,
    \begin{align*}
        \Cov(p,h)=\E[zh] &\geq\E[\abs{z}]-2\E[\abs{z}\one_{\domain(f)^c}]-2\E[\abs{z}\one_E]\\
        &\geq v-v/2-2\Pr[X\in E]\geq\frac{3v}{8}.
    \end{align*}
    Replacing each output $b$ by $2b-1$ gives an algorithm for $h$ using at most $D$ queries. Finally,applying \cref{lem:app-covariance} gives us the conclusion
    \begin{align*}
        \max_i\Inf_i(p) &\geq\frac{4\Cov(p,h)^2}{D^2}\geq\frac{9v^2}{16D^2}=\Omega\left(\frac{v^{12}}{d^{12}}\right).\qedhere
    \end{align*}
\end{proof}

\end{document}